\documentclass[onecolumn,floatfix]{revtex4}
\usepackage{color}
\usepackage{epsfig}
\usepackage{graphicx}
\usepackage{amsmath}

\newcommand {\be}  {
\begin{equation}
}
\newcommand {\ee}  {
\end{equation}
}
\newcommand {\bea} {
\begin{eqnarray}
}
\newcommand {\eea} {
\end{eqnarray}
}

\newcommand{\RN}[1]{%
  \textup{\uppercase\expandafter{\romannumeral#1}}%
}
\newcommand{\ignore}[1]{}

\usepackage{hyperref}

\begin{document}

\title{Geometry of triple junctions during grain boundary premelting\footnote{Uploaded with permission from APS as follows: M. Torabi Rad, G. Boussinot, and M. Apel, Physical Review Letters; Accepted, in Production (\href{https://journals.aps.org/prl/accepted/0907dY31S5f1f278887f0571d7a74e3ac98ad26d5}{Link to abstract}). }}

%
%

\author{M. Torabi Rad, G. Boussinot\footnote{First and second authors contributed equally.}\footnote{G.Boussinot@access-technology.de}, and M. Apel}
\affiliation{Access e.V., Intzestr. 5, 52072 Aachen, Germany}
\begin{abstract}

Grain Boundaries (GB) whose energy is larger than twice the energy of the solid/liquid interface exhibit the premelting phenomenon, for which an atomically thin liquid layer develops at temperatures slightly below the bulk melting temperature. Premelting can have a severe impact on the structural integrity of a polycrystalline material and on the mechanical high temperature properties, also in the context of crack formation during the very last stages of solidification. The triple junction between a dry GB and the two solid/liquid interfaces of a liquid layer propagating along the GB cannot be defined from macroscopic continuum properties and surface tension equilibria in terms of Young's law. We show how incorporating atomistic
scale physics using a disjoining potential regularizes the state of the triple junction and yields an equilibrium with a well-defined microscopic contact angle. We support this finding by dynamical simulations using a multi-phase field model with obstacle potential for both purely kinetic and diffusive conditions. Generally, our results should provide insights on the dynamics of GB phase transitions, of which the complex phenomena associated with liquid metal embrittlement are an example.

\end{abstract}

\maketitle


\emph{Introduction}.--- Triple junctions exist when three phases or grains meet. Their shape and especially contact angles, defined as the angles with which the interfaces intersect, are an interesting research subject in material science. These angles provide the boundary conditions that are the key in determining the morphology of an interface network, e.g. the grain boundary network in a polycrystal, as well as its temporal evolution, e.g. during coarsening. Young was the first to study contact angle selection on a macroscopic level \cite{Young}. According to his well-known law, the equilibrium contact angles are those that fulfill a force balance at the triple junction that involves only macroscopic surface tensions. The law is still widely used in numerous applications but disregards physical effects at the scale of the interatomic distance that may reveal crucial in some cases.\\

Physical effects originating from the interaction between interfaces when they are separated by only a few atomic distances can be modeled by adding a so-called disjoining potential to the free energy of the system. This potential, which can be either a monotonous function of the distance between interfaces $h$ or a function with one or multiple minima \cite{Mishin}, represents the work required to bring the interfaces from an infinite distance to the distance $h$; it is non-zero only when $h$ is in the order of interatomic distance. The negative of its derivative gives the disjoining pressure, a term commonly used in studying thin films \cite{Gennes}. For such films, the critical role of the disjoining potential in determining the contact angles at the junction is well known. For example, it is experimentally observed that in de-wetting of a disordered melt of diblock copolymer on a solid substrate, the existence of multiple minima in the disjoining potential results in a discrete set of equilibrium contact angles at the triple junction \cite{quantized}. It is also admitted that the disjoining potential plays a role in the stabilization of an atomically thin and strained Ge layer on a Si substrate \cite{SiGe}.  \\

When a polycrystalline material is at a temperature well below the bulk melting temperature $T_M$, the width of the grain boundaries (GBs) corresponds to a few atomic distances only. However, as the temperature increases towards $T_M$, GBs may transform into liquid-like, thermodynamically stable thin (i.e., nanoscale) films. Such an order-disorder transition is termed GB premelting \cite{NaturemetallicGBs} and may be observed not only at GBs but also on surface \cite{NatureSurface, NatureSurface2}. Due to its universal physics and important consequences in a wide range of applications, premelting has interested scientists from domains such as chemistry, physics, earth science \cite{rockpremelting}, fluid mechanics \cite{fluidmechanics}, and, interestingly, biology \cite{biology}. It has been modeled using different techniques such as molecular dynamics \cite{MD_1,MD_2}, phase-field-crystal \cite{phase_field_crystal}, and phase-field \cite{phase_field}. 
Premelting belongs to the class of surface phase transitions, for example between so-called GB complexions \cite{carter_PRB}.  Let us note that a similar situation is observed in the liquid film embrittlement of metals, e.g. by liquid Ga penetrating along Al GBs \cite{Ga}, for which the penetration speed depends crucially on the triple junction geometry. 
\\

\begin{figure}
  \centering
  \includegraphics[keepaspectratio, width=0.5\textwidth]{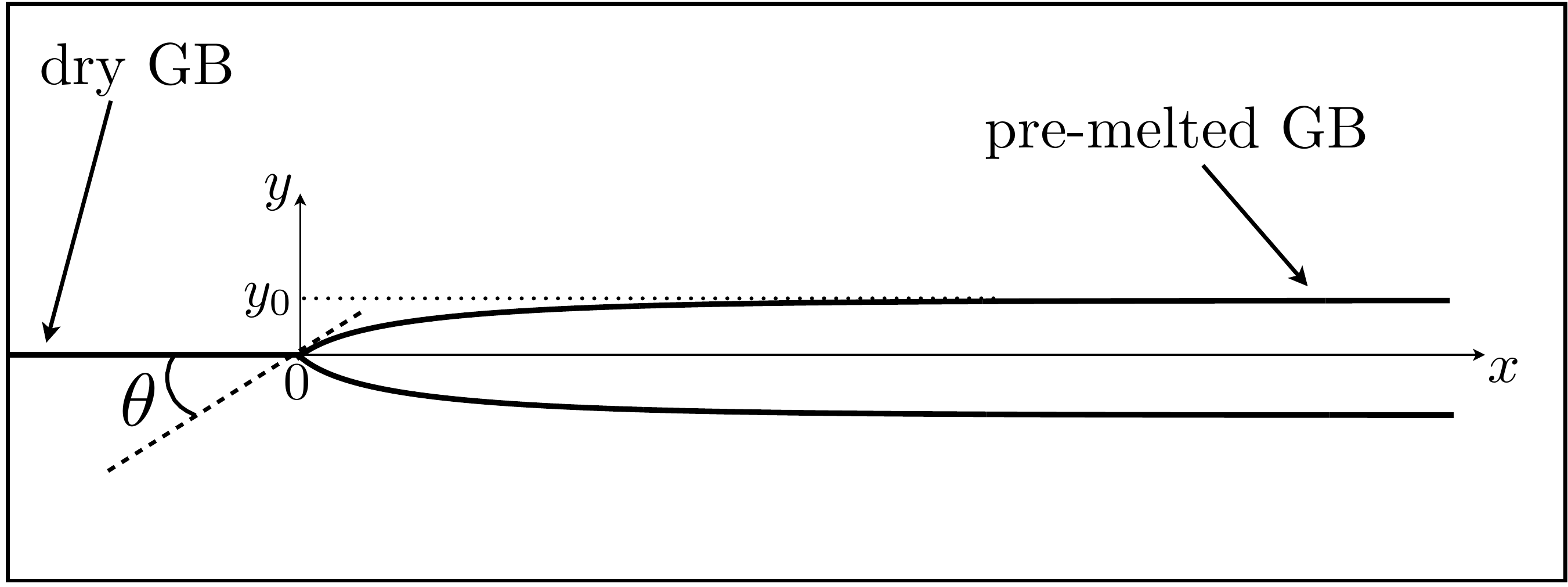}
  \caption{A schematic of the triple junction that forms when the pre-melted part of a GB meets its dry part} 
  \label{scheme}
\end{figure}

When premelting occurs in a polycrystalline material exposed to a temperature close to $T_M$, the liquid layer propagates along the GBs, and triple junctions form where the pre-melted part of the GBs meets their dry part. A schematic of such a junction is displayed in Fig. \ref{scheme}. The pre-melted part is bounded by two solid-liquid interfaces each having an energy $\sigma_{sl}$, and the energy of the dry GB is $\sigma_{gb}$. For such a junction, the contact angle $\theta$ verifies, according to Young's law:
\bea\label{Youngs_law}
2\cos(\theta) = \sigma^*
\eea 
where $\sigma^* = \sigma_{gb}/\sigma_{sl}$.
Pre-melting takes place when replacing the dry GB by two solid-liquid interfaces becomes energetically favorable, i.e. when $\sigma^* > 2$. Then, Eq. (\ref{Youngs_law}) does not have a solution. This indicates that for triple junctions that form during GB premelting, Young's law is not applicable. In this letter, we first show, using a sharp interface model, how introducing the disjoining potential into the energetics of the system regularizes the situation and yields an equilibrium with a well-defined microscopic contact angle, for which we give an analytical expression. We support this finding using multi-phase field model with obstacle potential
by an investigation of the dynamical state not only under purely kinetic conditions but also under diffusion controlled conditions. We finally conclude and discuss our results in the context of surface phase transitions.\\


\emph{Equilibrium microscopic contact angle}.--- We start studying the triple junction displayed in Fig. \ref{scheme} by considering a system that is in equilibrium at a temperature $T$ below the melting temperature $T_{M}$. The dimensionless temperature $\Delta$ is defined as
\bea
\Delta = c_P (T-T_M)/L 
\eea
where $c_P$ and $L$ are the specific and latent heats per unit volume, respectively; $\Delta$ represents the bulk free energy density difference between liquid and solid, and is negative in our range of interest. \\

In the sharp interface model, the solid-liquid interface is described by the function $y(x)$ with the triple junction at the origin, i.e. $y(0)=0$. The line $y=0$ is the axis of symmetry, and we solve for $x>0$ and $y>0$, i.e. for the upper interface in Fig. \ref{scheme}. We assume $\sigma^*-2  \ll 1$, allowing us to use a small slope approximation, i.e. $y'(x) \ll 1$. In this case, the free energy of the system coming from the interaction between the two solid/liquid interfaces may be written as an integral over $x$ of a local contribution, the disjoining potential, which is proportional to $\sigma_{gb}-2\sigma_{sl}$ and to a dimensionless function $f[y(x)]$. The latter describes the fact that the interface energy changes from $\sigma_{gb}$ when $y=0$, to $2\sigma_{sl}$ when $y \to \infty$. 
\\

In addition, the free energy includes two other terms. The first one is the bulk free energy increase due to the presence of the liquid phase below $T_M$, which is proportional to the entropy of fusion $L/T_M$, to the undercooling $T_M-T$ and to the amount of liquid $2y$ at position $x$. The second term represents the interface energy change $2 \sigma_{sl} (1+y'^2/2) - \sigma_{gb}$ linked to the substitution of the dry GB by two curved solid-liquid interfaces.
The free energy of the system then reads
\bea\label{free}
F[y(x)] = \int_{0}^\infty dx \left\{-  2y(x) \frac{L}{S} \Delta + \sigma_{sl} (\sigma^*-2) f[y(x)] + \sigma_{sl} \left( 2+ [y'(x)]^2  - \sigma^* \right)\right\}
\eea
where $S=c_P T_M/L$. \\

At equilibrium, the temperature and, therefore, $\Delta$ are homogeneous and $y(x)$ must satisfy $\delta F/\delta y(x) = 0$, i.e.
\bea\label{chempot_gen}
-2\Delta + (\sigma^*-2) d_0 \frac{df}{dy}[y(x)] - 2d_0 y''(x) = 0
\eea 
where $d_0 = S \sigma_{sl}/L$ is the capillary length.

Asymptotically far from the triple junction, the pre-melted GB is at equilibrium with a width $y_0$ such that $ d_0 \frac{df}{dy}(y_0) = 2\Delta/(\sigma^*-2)$, $y'(x \to \infty) = 0$, and $y''(x \to \infty) = 0$. Multiplying Eq. (\ref{chempot_gen}), which is satisfied at any position $x$, by $y'(x)$ and integrating over $x$ \cite{Spatschek} yields
\bea
\theta_0 = \left[2 \Delta \frac{ y_0}{d_0} + (\sigma^*-2) \{f(0)-f(y_0)\} \right]^{1/2}
\eea
where, $\theta_0 = y'(0)$ is the contact angle at the triple junction. 
Here, we find the equilibrium contact angle that derives from introducing the disjoining potential.
It thus corresponds to a {\it microscopic contact angle}, while at larger scale, the {\it macroscopic contact angle} vanishes ($y' \to 0$ when $x\to \infty$), in accordance with the condition of full wetting, i.e. $\sigma^*>2$. This is, to our best knowledge, the first derivation of the microscopic contact angle under a condition of full wetting of a GB. 

With $f(y) = \exp(-2y/d_w)$, for which $y_0 = (d_w/2) \ln \alpha$ where $\alpha = - (\sigma^*-2)(d_0/d_w)/\Delta$, one obtains
\bea\label{theta0}
\theta_0 = \left[(\sigma^*-2) \; \left( 1 - \frac{1+\ln \alpha}{\alpha}  \right) \right]^{1/2}.
\eea
We see that $\theta_0$ varies between $\sqrt{\sigma^*-2}$ and 0 within the pre-melting temperature range, i.e. for respectively $1/\alpha \to 0$ and $1/\alpha \to 1$. Thus, our small slope hypothesis indeed corresponds to the condition $\sigma^*-2 \ll 1$, and more precisely $\sqrt{\sigma^*-2} \ll 1$.
\begin{figure}
  \centering
  \includegraphics[keepaspectratio, width=0.7\textwidth]{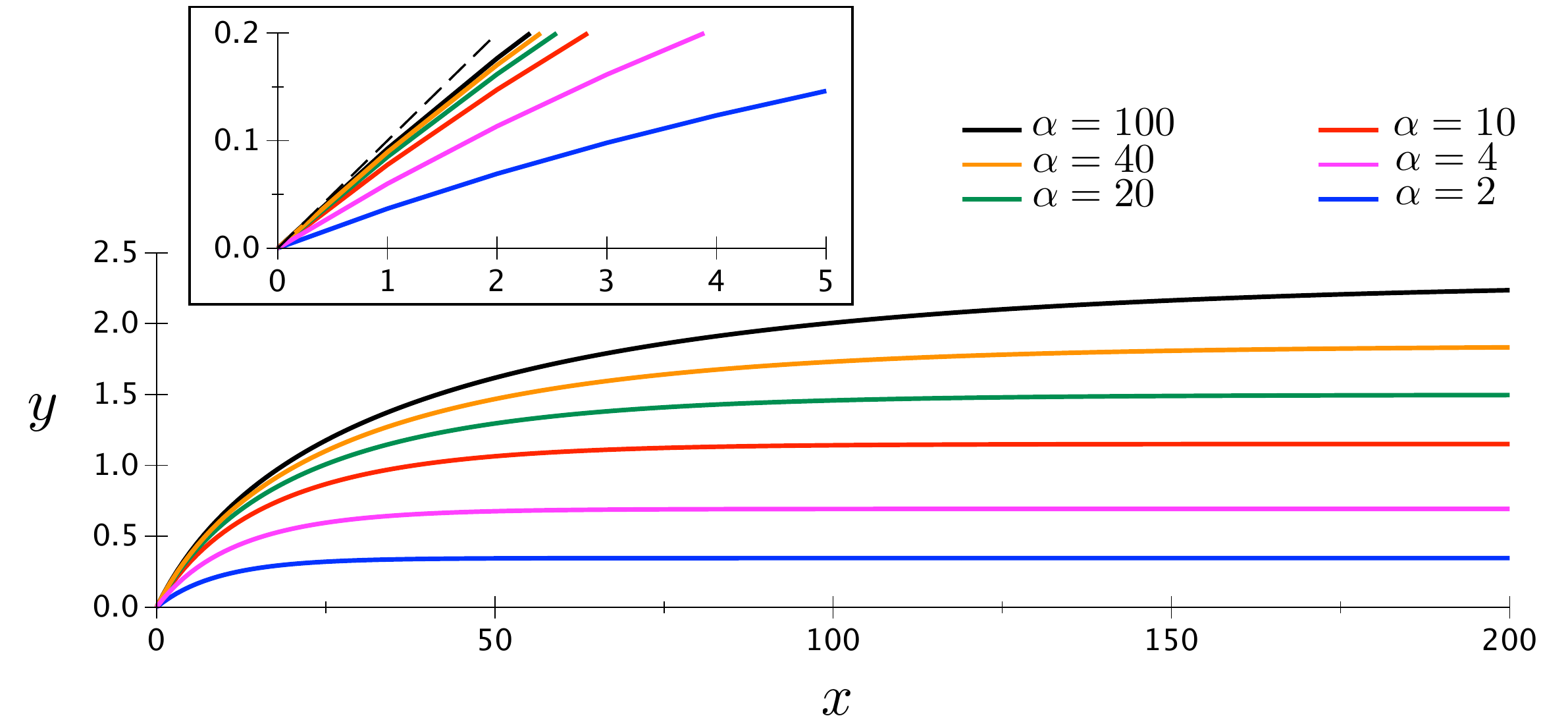}
  \caption{Shape of solid-liquid interface for different values of $\alpha$ in the case of the disjoining potential $f(y) = \exp(-2y/d_w)$. Here $\sigma^* = 2.01$ and $d_w/d_0=1$. Inset: focus on the region close to $x=0$ in which the $\alpha$-dependence of the microscopic contact angle and its convergence to $\sqrt{\sigma^*-2} = 0.1$ (dashed line) are apparent. } 
  \label{shapes}
\end{figure}
In Fig. (\ref{shapes}), we present the function $y(x)$ for different values of $\alpha$, for $\sigma^* = 2.01$ and for a ratio of microscopic lengths $d_w/d_0 = 1$. We observe the logarithmic increase of the equilibrium width $y_0$ with $\alpha$. In the inset, we present a close-up at the triple junction showing the convergence of $y'(0)=\theta_0 \to \sqrt{\sigma^*-2}$ (here 0.1, i.e. the dashed line) when $\alpha \to \infty$. 

\begin{figure}
  \centering
  \includegraphics[keepaspectratio, width=0.4\textwidth]{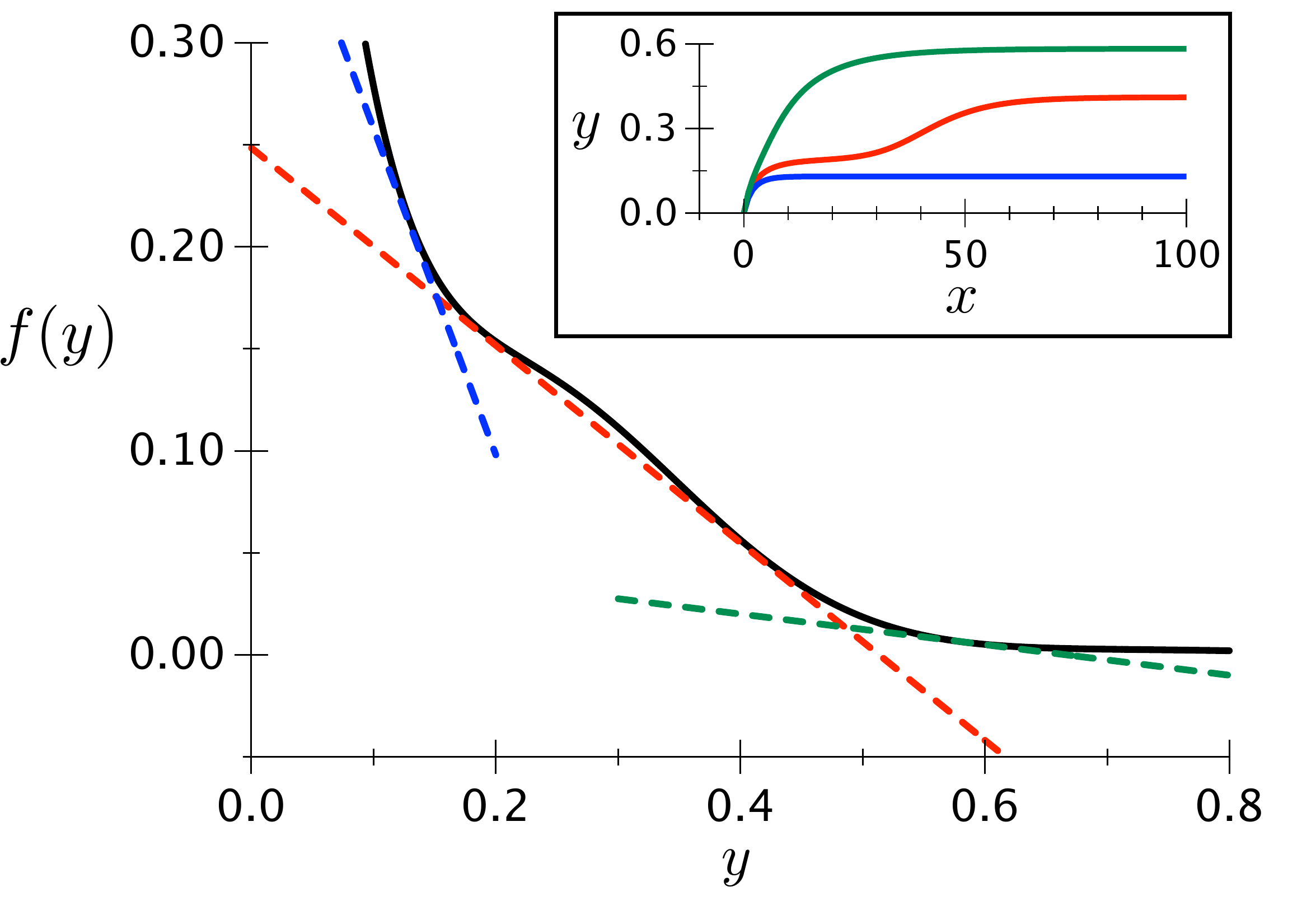}
  \caption{Non-convex disjoining potential (see text) with different scenarios depending on the slope of the dashed lines. Inset: corresponding equilibrium shape of the pre-melted layer. } 
  \label{gaal}
\end{figure}

In the case of a non-convex potential $f(y) = \exp \big(-2y/d_{w1} \big) \big(1 - \beta \sin 2y/d_{w2}) \big)$, a first-order GB phase transition occurs when $\Delta$ varies ("intermediate case" in Fig.~ 3 of Ref. \cite{Mishin}). As shown in Fig. \ref{gaal} (for $\sigma^*=2.01$, $d_{w1}/d_0 = 0.25$, $d_{w2}/d_0 = 0.16$ and $\beta=0.4$), $\Delta/(\sigma^*-2)=-0.8$ yields a low temperature equilibrium (blue) with a small layer width, while $\Delta/(\sigma^*-2)=-0.0375$ yields a high temperature equilibrium with a larger width (green). At $\Delta_c/(\sigma^*-2) \simeq -0.242$, a common tangent exists. Then, when $|\Delta|$ is larger than $|\Delta_c|$ the high-temperature phase is absent at equilibrium, and a coexistence of the two solutions occurs when $|\Delta|$ is slightly smaller than $|\Delta_c|$, as shown in the inset (red).   
Such a situation with multiple equilibria yields a diversity of GB phases in a polycrystalline material, as for example when few monolayer thick Ga, followed by a Ga liquid film, penetrates along GBs in polycrystalline Al  \cite{srolovitz, Ga}, or when several complexions are observed on the same GB in Ni-Bi \cite{ni_bi}.  
\\
\emph{Phase-field simulations}.--- Now, we investigate whether a phase-field model can reproduce the physics that was evidenced in the previous section. We use here the multi-phase field model with obstacle potentials implemented in the software MICRESS \cite{website}, which has proven its applicability in modeling pre-melting under equilibrium conditions.  Within this frame, the coupled evolution equations of $\phi_1, \phi_2$ and $\phi_3$ yield as an implicit result a convex disjoining potential, that obeys Eq. 46 in \cite{Bhogireddy} and differs from a simple exponential. Recently, we  applied this model to the problem of growth of the liquid layer along a dry GB \cite{our_paper}. The investigation of the structure of the triple junction during this growth is thus a natural extension of our previous study. \\

\begin{figure}
  \centering
  \includegraphics[keepaspectratio, width=\textwidth]{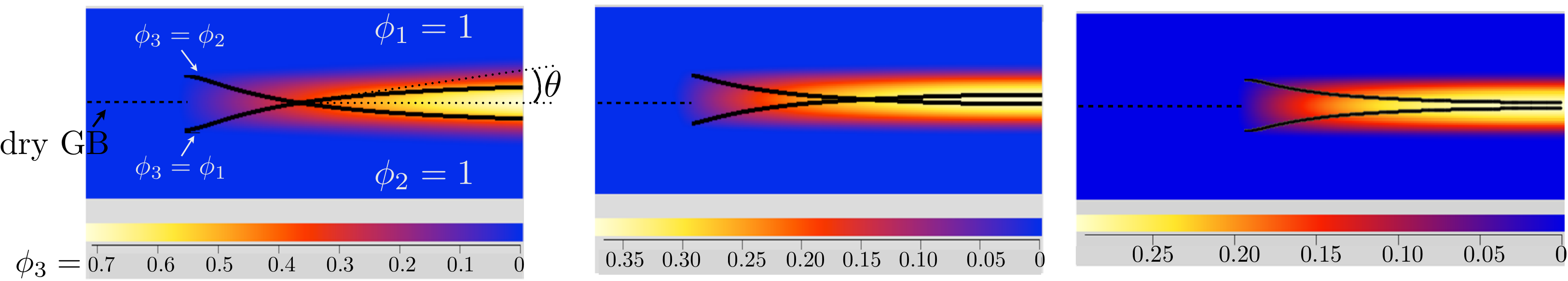}
  \caption{Color map of liquid phase-field $\phi_3$ along with iso-lines $\phi_3=\phi_2$ and $\phi_3=\phi_1$, whose intersection determines the triple junction position. Results are for $1/\alpha_{PF}=0.0469, 0.797, 0.992$ (left, center, right). In the latter case, because $1/\alpha_{PF} > 8/9$ the phase field computation does not lead to a defined triple junction. Note that the scale of the color bar is different between the left, center, and right plots.} 
  \label{profiles}
\end{figure}

The details of the phase field model are discussed elsewhere \cite{our_paper, Janin, MICRESS_paper}. As displayed in Fig. \ref{profiles}, grain 1 is assigned the phase-field $\phi_1$, meaning that $\phi_1=1$ in its bulk. Similarly, grain 2 is assigned the phase-field $\phi_2$.
The liquid phase is assigned the phase field $\phi_3$. The evolution equations obey a sum rule $\phi_1 + \phi_2 + \phi_3 = 1$, and for our pre-melting problem, the liquid is never in its bulk state, i.e. we have $\phi_3<1$. 
The triple junction is defined as the position where the two lines corresponding to $\phi_3 = \phi_1$ and $\phi_3 = \phi_2$ cross \cite{folch}, and the triple junction thus corresponds to an equality of the three phase fields $\phi_1 = \phi_2 = \phi_3 = 1/3$. The contact angle is then defined as the slope of these lines at the triple junction. \\

In our previous study \cite{our_paper}, we have analyzed in details the equilibrium corresponding to a pre-melted GB. The phase field equations are then one-dimensional, i.e. depend on a single spatial coordinate $x$, and they yield temperature-dependent phase fields distributions $\phi_1(x), \phi_2(x)$ and $\phi_3(x)$. The lengths are scaled by the interface width $\eta$, which corresponds to the length scale of the variations of the phase fields in case of a two-phase equilibrium, i.e. when only two phase fields vary while the other one vanishes identically. The liquid phase field $\phi_3$ reaches its maximum value at the center of the liquid layer, $x=0$. 
This maximum value $\phi_3(x=0)$ vanishes when a dimensionless temperature defined as
\bea
\Delta_{PF} =  (\tilde L/S) \Delta
\eea
with $\tilde L = L\eta/(4\sigma_{sl})$, reaches $-(\sigma^*-2)/2$. The latter quantity sets the maximum undercooling below which premelting is still possible. Recalling that, in the sharp interface model, $\alpha$ was defined as the ratio of the maximum undercooling for premelting and the actual undercooling, a corresponding quantity can be defined for the phase-field model as
\bea
\alpha_{PF} = - \frac{\sigma^*-2}{2 \Delta_{PF}} \;,
\eea
which also satisfies $0<1/\alpha_{PF}<1$. 
While the width of the liquid layer increases with $\alpha$ in the sharp-interface description, $\phi_3$ increases with $\alpha_{PF}$ in the phase-field model according to 
\bea\label{phi3}
\phi_3(0) \simeq \sqrt{1-1/\alpha_{PF}} 
\eea
when one assumes $\sigma^*-2 \ll 1$  (see Eq. 9 in \cite{our_paper} - pay attention to the definition of $\Delta$).\\

The objective of this section is to analyze whether the contact angles resulting from the phase field simulations reproduce, at least qualitatively since the disjoining potential is not a simple exponential in the phase field model, the temperature variation of the equilibrium contact angle predicted by  Eq. (\ref{theta0}). 
\\

Two different sets of phase field simulations were performed. First, a homogeneous and constant temperature field is prescribed and the dynamics is purely kinetically controlled. Second, diffusive conditions are assumed, for which the temperature field is numerically solved with a Dirichlet boundary condition far ahead of the triple junction and adiabatic boundary conditions elsewhere. In both cases, the liquid layer grows steadily along the dry GB. The characteristics of the produced pre-melted GB are set by the temperature far behind the triple junction, which is prescribed in kinetic conditions and results from energy conservation in diffusive conditions. In both sets of simulations, the grid step is $\eta/30$, $S=10$, $\tilde L = 37.5$. For the diffusive simulations, the diffusion coefficient $D$, expressed through the characteristic velocity $\mu$ associated with the phase field (phase field mobility)  is: $D / (d_0 \mu) = 1$. \\

As mentioned above, the location of the triple junction is determined by the intersection of the iso-lines $\phi_1=\phi_3$ and $\phi_2=\phi_3$, i.e. all the phase-fields take the same value of $1/3$ at the triple junction. Thus for $\phi_3(0) < 1/3$, i.e. for $1/\alpha_{PF} > 8/9$ (see Eq. (\ref{phi3})), the triple junction cannot be defined. This is  illustrated in Fig. \ref{profiles}, where  isolines $\phi_3=\phi_2$ and $\phi_3=\phi_1$ are superimposed on the color map of $\phi_3$ for three different values of $1/\alpha_{PF}$. It can be seen that, while the two iso-lines intersect for $1/\alpha_{PF} = 0.0469$ and 0.797, they do not for 0.992 and the position of the triple junction cannot be defined. \\

\begin{figure}
  \centering
  \includegraphics[keepaspectratio, width=0.75\textwidth]{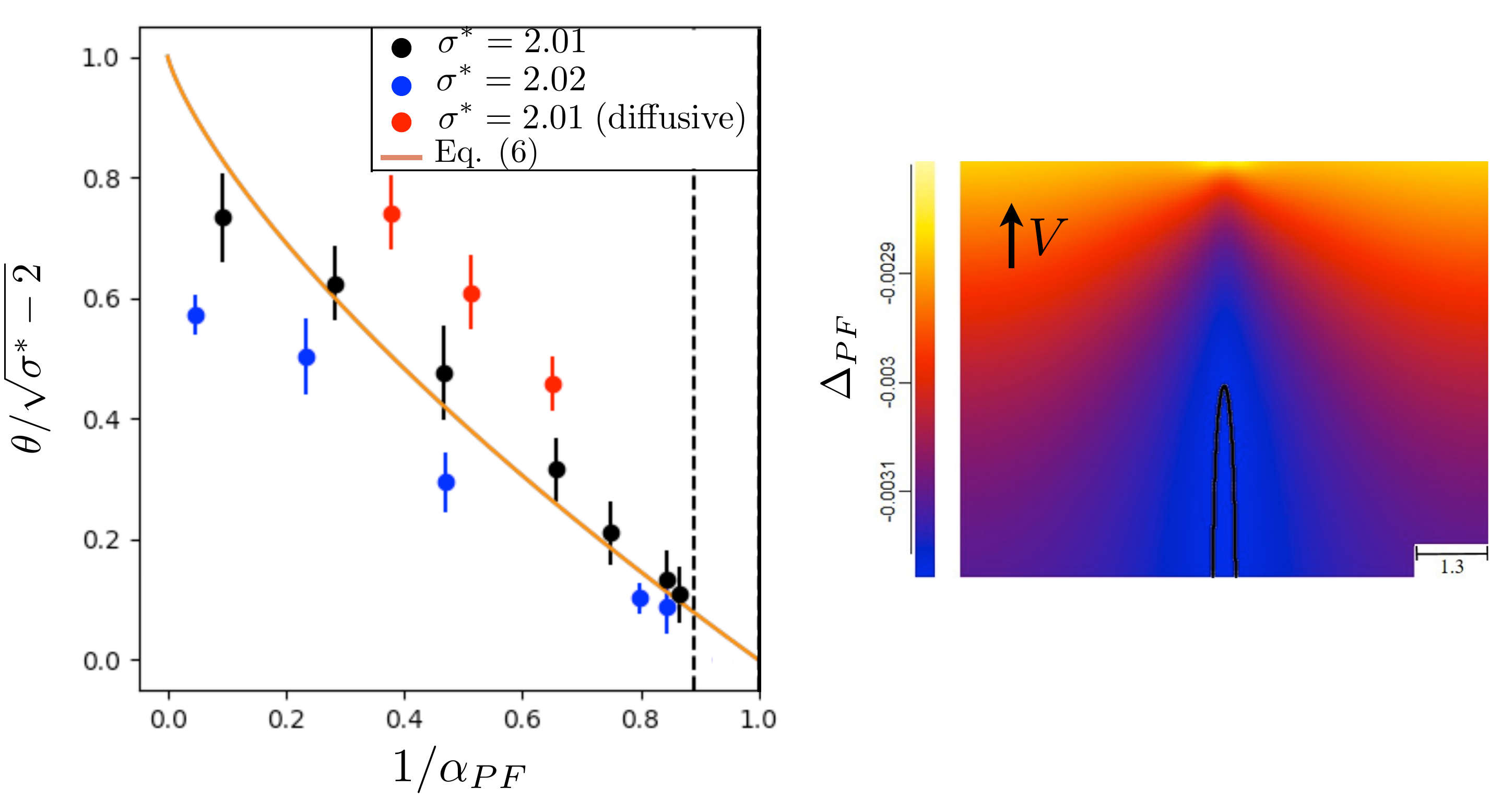}
  \caption{Left: Normalized microscopic contact angle measured in the phase field simulations for purely kinetic (blue and black markers) and diffusive (red markers) conditions as a function of the normalized undercooling $1/\alpha_{PF}$. The error bars correspond to the influence of the spatial discretization on the measurement. As a reference, the angle deriving from Eq. (\ref{theta0}) is plotted in brown solid line. The dashed vertical line corresponds to $1/\alpha_{PF} = 8/9$ (see text); Right: Diffusion field around the triple junction, located at the tip of the black curve, that represents $\phi_3 = 1/3$. The lengths are given in units of $\eta/(2\pi)$.} 
  \label{last_fig}
\end{figure}

In the left panel of Fig. \ref{last_fig}, the phase field results for the measured angle at the triple junction $\theta$ are presented as a function of $1/\alpha_{PF}$. Accordingly, no data point exists beyond the vertical dashed line, i.e. for $1/\alpha_{PF} > 8/9$. The angle is normalized by $\sqrt{\sigma^* - 2}$, and an error bar is provided corresponding to the effect of the spatial discretization on the measurement. The blue and black markers represent simulations under purely kinetic conditions for which $\alpha_{PF}$ is prescribed, respectively for $\sigma^* = 2.02$ and $2.01$. 
The red markers represent simulations with $\sigma^* = 2.01$ under diffusive conditions. In the right panel of Fig. \ref{last_fig}, we present the diffusion field in the neighborhood of the triple junction, with $1/\alpha_{PF} \simeq 0.65$ being measured at the tip of the black curve corresponding to $\phi_3 = 1/3$. The inhomogeneity of $\Delta_{PF}$ on the relevant length scales ($\eta$ in the $y$-direction and $\eta/\sqrt{\sigma^*-2}$ in the $x$-direction) is sufficiently small (not exceeding 7-8\% of its value in any simulation) for excluding its influence on the difference of measured angle whether in purely kinetic or diffusive simulations (black and red markers). Instead, we attribute this difference to finite velocity effects. Indeed, here, for $1/\alpha_{PF} = 0.38, 0.51$ and 0.65, the selected velocities are $Vd_0/D = 1.375, 0.975$ and 0.6875, respectively, and the ratio between the growth velocity and the diffusion velocity is of order unity, while vanishing velocity effects would correspond to  $Vd_0/D \ll 1$.

In addition, the analytical expression for the equilibrium contact angle in Eq. (\ref{theta0}) is given as a reference with a brown solid line in the left panel. A similar trend is clearly observed in the simulations and the analytics.  The scaling $\theta \sim \sqrt{\sigma^*-2}$ seems fully appropriate in view of the proximity of the different sets of data points and of the analytical curve, although the condition $\sqrt{\sigma^* - 2} \ll 1$ is actually relatively weakly fulfilled.
\\

\emph{Conclusion}.--- We have studied, using sharp interface and phase field models, the geometry of a triple junction under pre-melting conditions, i.e. when an atomically thin liquid layer propagates along a dry grain boundary (GB) below $T_M$. While Young's law fails in this case, the disjoining potential plays a major role and provides a solution for the microscopic contact angle and the shape of the liquid layer. 
Our results can be interpreted in terms of surface phase transitions, between the "dry" phase and one or more (when the disjoining potential is non-convex) pre-melted phases.  
The integrand in Eq. (\ref{free}) varies along the $x$-axis between 0 in the dry phase ($x<0$) and a negative value when $x \to \infty$, with a maximum proportional to $\theta^2 \sim \sigma^*-2$ at the triple junction. Accordingly, the  transition takes place on a length of order $d_w/\theta \sim d_w/\sqrt{\sigma^* - 2}$.
The phase field simulations reproduce faithfully the analytical results for the contact angle, thus opening the way for the simulation of analytically non-tractable more complex scenarios.  While the phase field model used here generates a convex disjoining potential, an interesting future development of the model would concern non-convex potentials.
Our work contributes to the understanding of pre-melting, in particular in systems exhibiting liquid-metal-embrittlement such as Ga-Al \cite{Ga} or Ni-Bi \cite{ni_bi}, and of transitions between GB complexions such as the one reported in Ref. \cite{domino_pearl}. It may therefore open new ways for the engineering of transport properties along interfaces in materials. 

\section{Acknowledgements}

This study was funded by the Deutsche Forschungsgemeinschaft (DFG, German Research Foundation) under grant number AP196/17-1.

\end{document}